\documentclass{emulateapj}
\usepackage{graphicx}
\usepackage{amsmath}
\usepackage[varg]{txfonts}
\newcommand{\rmxaa}{RMxAA}
\begin{document}
\title{The Influence of Radial Stellar Migration on the Chemical Evolution of the Milky Way}

\author{Yue Wang\altaffilmark{1,2},Gang Zhao\altaffilmark{1}}
\altaffiltext{1}{Key Laboratory of Optical Astronomy, National Astronomical
Observatories, Chinese Academy of Science, Beijing 100012, China; gzhao@nao.cas.cn}
\altaffiltext{2}{University of Chinese Academy of Sciences, Beijing 100049, China}

\begin{abstract}
{Stellar migration is an important dynamical process in Galactic disk. Here we model the radial stellar migration in the Galactic disk with an analytical method, then add it to detailed Galactic chemical evolution model to study the influence of radial stellar migration on the chemical evolution of the Milky Way, especially for the abundance gradients. We found that the radial stellar migration in the Galactic disk can make the profile of the G-dwarf metallicity distribution of the solar neighborhood taller and narrower, thus it becomes another solution to the ``G-dwarf problem''. It can also scatter the age-metallicity relation. However, after the migration, the abundance distributions along the Galactic radius don't change much, namely the abundance gradients would not be flattened by the radial stellar migration, which is different from the predictions of many theoretical works. But it can flatten the radial gradients of the mean chemical abundance of stars, and older stars possess flatter abundance gradients than younger stars. The most significant effect of the radial stellar migration on the chemical abundance is that at a position it scatters the abundance of stars there from a relatively concentrated value to a range.}
\end{abstract}


%

\section{Introduction}
The evolution of the Milky Way is clearly manifested by its structure formation and the enrichment of chemical elements heavier than those stemming from the primordial nucleosynthesis, and the latter resulting from the consumption of gas to star formation. Gas collapses to form stars, and various chemical elements are created by nucleosynthesis processes in stars. Through several ways, such as mass loss and supernova explosion (during which heavier elements can also be produced), the newly synthesized nuclear products go back into the interstellar medium (ISM). Then the gas containing newly produced material forms stars of next generation, and the processes above will repeat again and again. The cycle between ISM and stars continuously enrich the Milky Way chemically on both the variety and the amount of chemical elements. Thus the dynamical behaviors of gas and stars would significantly influence the Galactic chemical evolution and contribute to form the present chemical features of the Milky Way.

To understand how the Galaxy formed and have evolved on both structure and chemical composition, models of the chemical evolution of the Milky Way play an important role in the research. From the earliest closed-box Simple Model to the open models with gas infall from outside of the Galaxy \citep{MG86, MF89, C97, P98}, and to the models including radial gas flow \citep[][the first one doesn't relax IRA]{LF85,PC00, SM11}, the chemical evolution model of the Milky Way has been developing all the time. Up to now the effects of gas part on the Galactic chemical evolution has been studied in detail.

However, the topics of how stars evolve dynamically in galactic disks and its effects on galactic chemical evolution have received less attention in the aforementioned works. One main reason resides that it has been believed for decades that stars would remain at roughly their birth radii throughout their lives. Thus, according to the mechanism of chemical enrichment, there would be a perfect relation between the age and metallicity of stars (the so-called Age-Metallicity Relation, AMR) at a certain position. For example, in the solar neighborhood the metallicity of stars decreases with the increasing of their age. This relation has been adopted by many chemical evolution models in the last decades of years \citep[e.g.][]{T75, MF89, C96, C97, BP99}. As a consequence, the influences of the dynamical behaviors of stars on the Galactic chemical evolution have not been paid enough attention.

However, as the enrichment of observational data and the development of theoretical studies in recent years, the classic picture has been replaced by new knowledge. Observationally, the tight age-metallicity relation has not been found in recent data. On the contrary the distribution of the age and metallicity of stars in the solar neighborhood is flat together with large scatter \citep{E93, N04}, and the scatter would increase with the increasing age \citep{H08, CGS}. It suggests that stars born at other regions with different age and metallicities could appear in the solar neighborhood through migration. Besides, the metallicity distribution function (MDF) in the solar neighborhood, which acts as a constraint of the chemical evolution model, obtained by \citet{CGS} shows that old stars have a considerably broader distribution than young stars while the peak always remains around the solar value, suggesting that the wings of MDF could mostly comprise stars born at various Galactocentric radii and migrated to the current position over different timescales. Theoretically, \citet{SB09a} has distinguished two drivers of radial migration: by scattering at an orbital resonance or by non-resonant scattering by a molecular cloud, the angular momentum of a star can be changed so that the star's guiding-center radius changes and the star's entire orbit moves inwards or outwards depending on whether its angular momentum is lost or gained, or the angular momentum of a star can be maintained while the star's epicycle amplitude increases so that the star contributes to the density over a wider range of radii. They call them churning and blurring respectively. Besides, \citet{MF10} identified another radial migration mechanism in barred galaxies, namely the resonance overlap of the bar and spiral structure induces a nonlinear response leading to a strong redistribution of angular momentum in the disk. As a consequence, besides gas flows, the radial stellar migration is another important component that should be considered in the Galactic chemical evolution model.

When the radial stellar migration is included in chemical evolution models, the classical tight correlation between age and metallicity is relaxed, and this relation becomes effectively an additional constraint independent from the metallicity distribution. Abundance gradients are also very important constraints for the Galactic chemical evolution model. As suggested by \citet{SB02} and \citet{SB09a}, since the migration of stars makes it possible to transport materials to anyplace along the disk no matter where they originally located, the mixture of chemical elements over large range of radius would reduce the differences of abundances between different regions along the disk and flatten the abundance gradients.

This paper is to study the influence of radial stellar migration in the disk on the chemical evolution of the Milky Way and the degree to which it could affect the formation of some Galactic typical chemical features. We organize it as follows: in Section 2 we describe the reference models used in this work; the description of the radial stellar migration is presented in Section 3; Section 4 and 5 show the stellar ejecta and the observational data adopted here, respectively; in Section 6 we report and discuss our results; and finally we draw our main conclusions in Section 7.


\section{Galactic chemical evolution model}
The model adopted here is based on the model of \citet{PC00}, which is an open model where the disk forms gradually by accretion of protogalactic gas and radial gas flow exists. The disk is divided into N concentric rings or shells, each of which is of 1~kpc wide; and in each ring the gaseous component and its chemical abundances evolve due to:
1. Star formation which locks up gas into stars;
2. Infall of primordial protogalactic gas;
3. The ejecta of stars which were born in the ring or migrated from other rings, bringing the enriched material back to the interstellar medium;
4. Gas exchange with the neighboring rings as a result of radial gas flows.

The basic equation describing the variation of gaseous component of the k-th ring is:
\begin{equation}
\begin{split}
{\dot\sigma}_{i}(r,t)&=-~\psi(r,t)X_{i}(r,t) \\
\smallskip
& +\sum_{r}\int\limits^{M_{Bm}}_{M_{L}}\psi(r,t-\tau_{m})R_{mi}(t-\tau_{m})\,\phi(m)\,f_{S\!M}(\tau_{m})~dm \\
& +\sum_{r}A\int\limits^{M_{BM}}_{M_{Bm}}\frac{\phi(m_{B})}{m_{B}}  \\
& ~~\times \left[~\int\limits_{\mu_{m}}^{0.5}f(\mu)~\psi(r,t-\tau_{m_{1}})R_{m_{1}i}(t-\tau_{m_{1}})\,m_{1}\,f_{S\!M}(\tau_{m_{1}})~d\mu \right. \\
&\qquad + \left. \int\limits_{\mu_{m}}^{0.5}f(\mu)~\psi(r,t-\tau_{m_{2}})E_{S\!N\!eI\!ai}\,f_{S\!M}(\tau_{m_{2}})~d\mu\right]~dm_{B} \\
&+\sum_{r}(1-A)\\
&\quad \times\int\limits^{M_{BM}}_{M_{Bm}}\psi(r,t-\tau_{m_{B}})R_{m_{B}i}(t-\tau_{m_{B}})~\phi(m_{B})\,f_{S\!M}(\tau_{m_{B}})~dm_{B} \\
&+\sum_{r}\int\limits^{M_{U}}_{M_{BM}}\psi(r,t-\tau_{m})R_{mi}(t-\tau_{m})~\phi(m)\,f_{S\!M}(\tau_{m})~dm \\
\smallskip
&+\left(X_{i}\right)_{inf}~{\dot\sigma}_{inf}(r,t)~+~\left[{\dot\sigma}_{i}(r,t)\right]_{gf}~,
\end{split}\label{evol}
\end{equation}
where $\sigma_{i}(r,t)$ is the surface gas density of element i at radius r at time t, $X_{i}(r,t)=\sigma_{i}(r,t)/\sigma_{gas}(r,t)$ is the abundance by mass of element i, $\psi(r,t)$ is the star formation rate at radius r at time t, $\phi(m)$ is the initial mass function, $R_{mi}(t)$ is the mass fraction of a star with mass m ejected into the ISM in the form of element i, and $E_{S\!N\!eI\!ai}$ is the mass contribution of a type Ia supernova explosion for element i. The first term represents the reduction caused by star formation. The following four integral terms describe the contributions of stars in different mass ranges. The summations represent the effects of stellar radial migration from all the considered radii, and $f_{S\!M}$ is the distribution function of radius describing the stellar migration which will be shown in detail in Section 3. The last two terms represent the contributions of gas accretion and radial gas flow, respectively.

\subsection{Star formation rate}
The star formation rate (SFR) $\psi(r,t)$ adopted here is of the form in \citet{TA75},
\begin{equation}\label{SFR}
\psi(r,t)=\nu\left[\frac{\sigma(r,t)~\sigma_{g}(r,t)}{\tilde{\sigma}^{2}(\tilde{r},t)}\right]^{(k-1)}\sigma_{g}(r,t)
\end{equation}
where $\nu$ is the efficiency of the star formation rate, $\sigma_{g}(r,t)$ is the surface mass density of gas, $\sigma (r,t)$ is the total surface mass density, $\tilde{\sigma}(\tilde{r},t)$ is the total surface mass density at a particular Galactocentric distance $\tilde{r}$ which is taken here to be $R_\odot=8.5~kpc$.

\subsection{Initial mass function}
The initial mass function (IMF) we adopt is assumed to be constant in space and time, and the prescription for it are taken from \citet{S86}:
\begin{equation}\label{IMF}
\phi(m)dm=cm^{-\mu}dm
\end{equation}
where $\mu=1.35$ for $m < 2~M_{\odot}$ and $\mu=1.7$ for $m \ge 2~M_{\odot}$. Since the majority of the chemical enrichment is due to stars of $m \ge 1~M_{\odot}$, it is meaningful to fix the fraction $\zeta$ of the total stellar mass distributed in stars above $1~M_{\odot}$. Thus the lowest stellar mass considered here, $M_{l}$ , is also fixed according to the normalization condition.
\begin{equation*}
\int\limits^{M_{u}}_{M_{l}}~\phi(m)dm=1
\end{equation*}
\begin{equation*}
\int\limits^{1M_{\odot}}_{M_{l}}~\phi(m)dm+\zeta=1
\end{equation*}
We adopt the upper mass limit, $M_{u}$, to be $120~M_{\odot}$.

\subsection{Gas accretion}
An infall rate of primordial protogalactic gas decreases exponentially in time with a timescale $\tau(r)$:
\begin{equation}\label{infall}
{\dot\sigma}_{inf}(r,t)= A(r)~e^{-\frac{t}{\tau(r)}},
\end{equation}
where $\tau(r)$ is chosen for two conditions: be constant at all radius and be variable with Galactic radius under an inside-out disk formation scenario. $A(r)$ is obtained by reproducing the current total surface mass density distribution, and the present surface mass density of the disk is exponential with the scale length $R_{D}=3.5~kpc$ and equals $54~M_{\odot}~pc^{-2}$ at the solar radius, and $17~M_{\odot}~pc^{-2}$ was adopted for the halo \citep{C01}.

\subsection{Stellar lifetime}
To drop the ``instantaneous recycling approximation (IRA)'', the finite lifetimes of stars of different masses must be considered. We adopt the metallicity-dependent stellar lifetimes from \citet{P98} (see their Table 14), in which the lifetimes are calculated as the sum of the H-burning and He-burning timescales of the stellar tracks of Padua library. In our models, for each mass M and metallicity Z we calculate the corresponding lifetime $\tau_{MZ}$ by interpolating within the logarithmic relation $\log(M)-\log(\tau)$ for the tabulated metallicities, and then by interpolating with respect to Z.

\subsection{Type Ia supernova}
Type Ia supernovas (SNe Ia) are important contributors of heavy elements, especially iron. Here we adopted the single degenerate model for the progenitors of SNe Ia, namely a degenerate C-O WD plus a red giant or main-sequence companion. The rate of SNe Ia is:
\begin{equation}
R_{S\!N\!eI\!a}(t)=A\int\limits^{M_{BM}}_{M_{Bm}}\phi(m_{B})\left[
  \int\limits^{0.5}_{\mu_{m}}f(\mu)~\psi(t-\tau_{m_{2}})~d\mu \right]~dm_{B}
\label{SNIa}
\end{equation}
where A is a free parameter representing the fraction in the IMF of binary systems with the right properties to give rise to SNe Ia. $M_{BM}$ and $M_{Bm}$ are the upper and lower limit respectively for the total mass of a binary system able to produce a SNe Ia. As we use the chemical ejecta of stars by models with convective overshoot (see Section 4), the corresponding upper limit to the mass of the primary is $6~M_{\odot}$, so $M_{BM}=12~M_{\odot}$,  while $M_{Bm}\sim3~M_{\odot}$ is adopted generally \citep{P98}. The function $f(\mu)$ describes the distribution of the mass ratio of the secondary $(\mu=M_{2}/M_{B})$ of the binary systems, and $f(\mu)=2^{1+\gamma}(1+\gamma)~\mu^{\gamma}$ with $\gamma=2$ as adopted by Greggio \& Renzini (1983a).

In this paper, we assume the ejecta of SNe Ia are released in two steps as adopted in \citet{P98}: at the end of the primary's life, it expels its products just like a single star, while after the period of the lifetime of the companion the supernova explodes. So the contributions of SNe Ia are calculated in two parts:
\begin{equation}
\begin{split}
\left[\dot{\sigma_{i}}(t)\right]_{Ia}&=A\int\limits^{M_{BM}}_{M_{Bm}}\frac{\phi(m_{B})}{m_{B}} \\
&\quad \times\left[\int\limits_{\mu_{m}}^{0.5}f(\mu)~\psi(t-\tau_{m_{1}})\,R_{m_{1}i}(t-\tau_{m_{2}})\,m_{1}\,~d\mu \right. \\
&\qquad\quad + \left. \int\limits_{\mu_{m}}^{0.5}f(\mu)~\psi(t-\tau_{m_{2}})\,E_{S\!NeIai}~d\mu\right]~dm_{B}
\end{split}
\label{SNIa-ejecta}
\end{equation}
where the first term in the square bracket represents the primary's contribution and the second term represents the contribution of the SNe Ia explosion.

\subsection{Radial gas flow}
Radial gas flow in Galactic disk can be driven by several mechanisms as introduced in \citet{LF85}, such as the difference between the angular momentum of the infalling gas and that of the circular motions in the disk, the viscosity in the gas layer, and the interactions between gas and the spiral structure. As the radial gas flows are expected to be inflows over most of the disk \citep{PC00}, we only consider inflows in our model, and we describe the gas inflows following the prescriptions in \citet{SM11}.

Let the k-th ring be represented by the galactocentric radius $r_{k}$ with its inner and outer edge being $r_{k-\frac{1}{2}}$ and $r_{k+\frac{1}{2}}$ $(r_{k-\frac{1}{2}}=(r_{k-1}+r_{k})/2$ , $r_{k+\frac{1}{2}}=(r_{k}+r_{k+1})/2)$. The velocities of the gas flowing through these edges are $v_{k-\frac{1}{2}}$ and $v_{k+\frac{1}{2}}$ respectively with positive outward and negative inward. The surface gas density of the k-th ring is changed due to gas flows with a flux $F(r)$ in the following way:
\begin{equation}
\label{gasflow}
\left[ \frac{d \sigma_{g}(r_{k},t)}{d t} \right]_{gf} =
   - \frac{1}{\pi ~\left( r^2_{k+\frac{1}{2}} - r^2_{k-\frac{1}{2}} \right) }
   \left[ F(r_{k+\frac{1}{2}}) - F(r_{k-\frac{1}{2}}) \right]
\end{equation}
where the gas flow at $r_{k+\frac{1}{2}}$ can be written as:
\begin{equation*}
\label{flux3}
F(r_{k+\frac{1}{2}}) = 2~\pi~r_{k+\frac{1}{2}} \, v_{k+\frac{1}{2}} \left[\sigma_{g}(k+1) \right].
\end{equation*}
Then we get:
\begin{equation}
\left[ \frac{d}{dt} \sigma_{i}(r_{k},t) \right]_{gf} = -\, \beta_{k} \,
\sigma_{i}(r_{k},t) + \gamma_{k} \, \sigma_{i}(r_{k+1},t),
\end{equation}
where $\beta_{k}$ and $\gamma_{k}$ are :
\begin{align*}
\beta_k &=  - \, \frac{2}{r_k + \frac {r_{k-1} + r_{k+1}}{2}}
	     \left[ v_{k-\frac{1}{2}}
	    \frac{r_{k-1}+r_k}{r_{k+1}-r_{k-1}}  \right]  \\
\gamma_k &=  - \, \frac{2}{r_k + \frac {r_{k-1} + r_{k+1}}{2}}
	     \left[ v_{k+\frac{1}{2}}
	     \frac{r_k+r_{k+1}}{r_{k+1}-r_{k-1}} \right].
\end{align*}

We assume there are gas flows from the outer parts of the disk where is beyond $r=18~kpc$, although there is no star formation in our model. We calculate the amount of gas flowing from the ring at $r=19~kpc$ to the ring at $r=18~kpc$ assuming that the surface gas density of the ring at $r=19~kpc$ evolves just like other rings which have star formation but its chemical composition stays primordial. This method may be not reasonable enough but it can easily help to reproduce the observational distribution of the gas along the radius at outer disk. Besides, the purpose of this paper is to study the effects of radial stellar migration, so we haven't spent much energy on the gas flow part. More detailed studies about the gas flow can be found in previous papers \citep[e.g.][]{PC00, SM11}.

The velocity of the gas flows adopted here is variable with radius as according to our tests, variable velocity is more helpful to reconstruct the radial abundance gradients along the Galactic disk, and is also more reasonable.

\begin{table*}
\begin{center}
\caption{Main parameters of the models.}
\begin{tabular}{ccccccc}
\hline \hline
\noalign{\smallskip}
 Models & Stellar migration & $\tau_D $ & $\zeta$ & $A$ & $\nu$ & $v_{gf}(r)$  \\
        &                   &  (Gyr)    &         &     & $(Gyr^{-1})$ & $(km~s^{-1})$   \\
\noalign{\smallskip}
\hline
\noalign{\smallskip}
GCM1 & No & 3 & 0.25 & 0.07 & 0.55 & $-0.08r-0.2, r < 10~kpc$  \\
\noalign{\smallskip}
\cline{1-6}
\noalign{\smallskip}
GCM1N(e) & Normal & 3 & 0.25 & 0.07 & 0.55 & $-0.2r+1.0, 10~kpc\le r < 15~kpc$  \\
\noalign{\smallskip}
\cline{1-6}
\noalign{\smallskip}
GCM1G(e) & Gama & 3 & 0.25 & 0.07 & 0.55 & $-1/3r+3.0, r\ge15~kpc$  \\
\noalign{\smallskip}
\hline
\noalign{\smallskip}
GCM2 & No & $0.88r-0.9$ & 0.3 & 0.09 & 0.9 & $-0.16r+0.6, r < 10~kpc$  \\
\noalign{\smallskip}
\cline{1-6}
\noalign{\smallskip}
GCM2N(e) & Normal & $0.88r-0.9$ & 0.3 & 0.09 & 0.9 & $-0.2r+1.0, 10~kpc\le r < 15~kpc$  \\
\noalign{\smallskip}
\cline{1-6}
\noalign{\smallskip}
GCM2G(e) & Gama & $0.88r-0.9$ & 0.3 & 0.09 & 0.9 & $-1/3r+3.0, r\ge15~kpc$  \\
\noalign{\smallskip}
\hline
\end{tabular}
\label{models}
\tablecomments{The `(e)' at the end of the name of a model including radial stellar migration indicates that the values of the parameters on the right are adopted for both of the models adopting standard and extreme condition respectively for stellar migration, eg. GCM1N and GCM1Ne. $\tau_D $ is the disk formation time scale, $\zeta$ is the fraction of the total stellar mass distributed in stars above $1~M_{\odot}$, $A$ is the parameter representing the fraction in the IMF of binary systems with the right properties to give rise to SNe Ia, $\nu$ is the efficiency of the star formation rate, $v_{gf}(r)$ is the velocity of radial gas flow.}
\end{center}
\end{table*}

~\\

\begin{figure}
\centering
\includegraphics[width=3.4in]{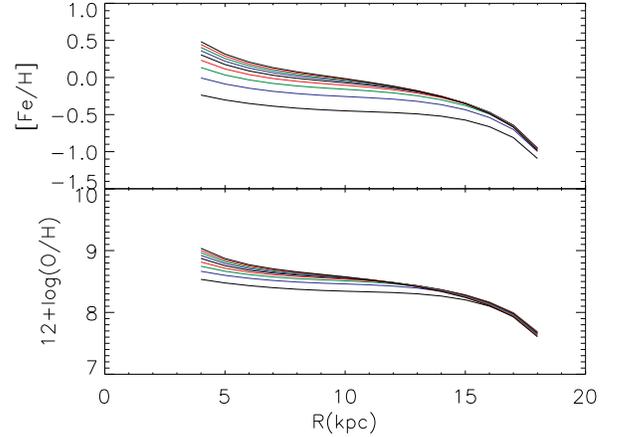}
\caption{Galactic radial iron (top) and oxygen (bottom) abundance distribution calculated by model GCM2, the lines from the bottom to the top represent the abundance distributions from $t=4~Gyr$ to $t=20~Gyr$, respectively.}
\label{Gasflow-TimeGradients}
\end{figure}

As the effects of all the factors above have already been studied in detail in many previous papers \citep[e.g.][]{P98, PC00, SM11}, and the purpose of this paper is to find out the influence of the radial stellar migration on the Galactic chemical evolution, we just adopt some typical prescriptions for the parts except stellar migration in the models as the bases to study stellar migration. The main parameters adopted for all the models are listed in Table \ref{models}. Two kinds of Galactic disk formation scenarios are adopted: 1. the whole disk formed simultaneously with a constant disk formation time scale for all the Galactocentric radii, and these models are titled with `GCM1'; 2. the disk formed following an inside-out formation scenario with the disk formation time scale increasing with radius, and these models are titled with `GCM2'. The `(e)' at the end of the name of a model including radial stellar migration indicates that the values of the parameters on the right are adopted for both of the models adopting standard and extreme condition respectively for stellar migration (see Section 3), eg. GCM1N and GCM1Ne. GCM1 and GCM2 are the basic models without radial stellar migration. Figure \ref{Gasflow-TimeGradients} shows the time-dependent radial iron and oxygen abundance distributions calculated by model GCM2 as an example.

\section{Description of the radial stellar migration}

We describe the radial stellar migration in an analytical way, using different distribution functions to model the probability that stars, born at a given radius, appear at different radii after migrating for a period of time. This method doesn't refer to any single star but describes the radius distribution of stars quantitatively in a statistical way. Different from single stars, the migration of binaries may be more complex, but as the binaries always orbit around each other closely, we just assume the binary systems migrate like single stars. Thus the progenitors of SNe Ia would migrate radially, the effects of which are considered in our calculation of the chemical contributions of SNe Ia.

\begin{figure*}
\centering
\includegraphics[width=16.0cm, angle=0]{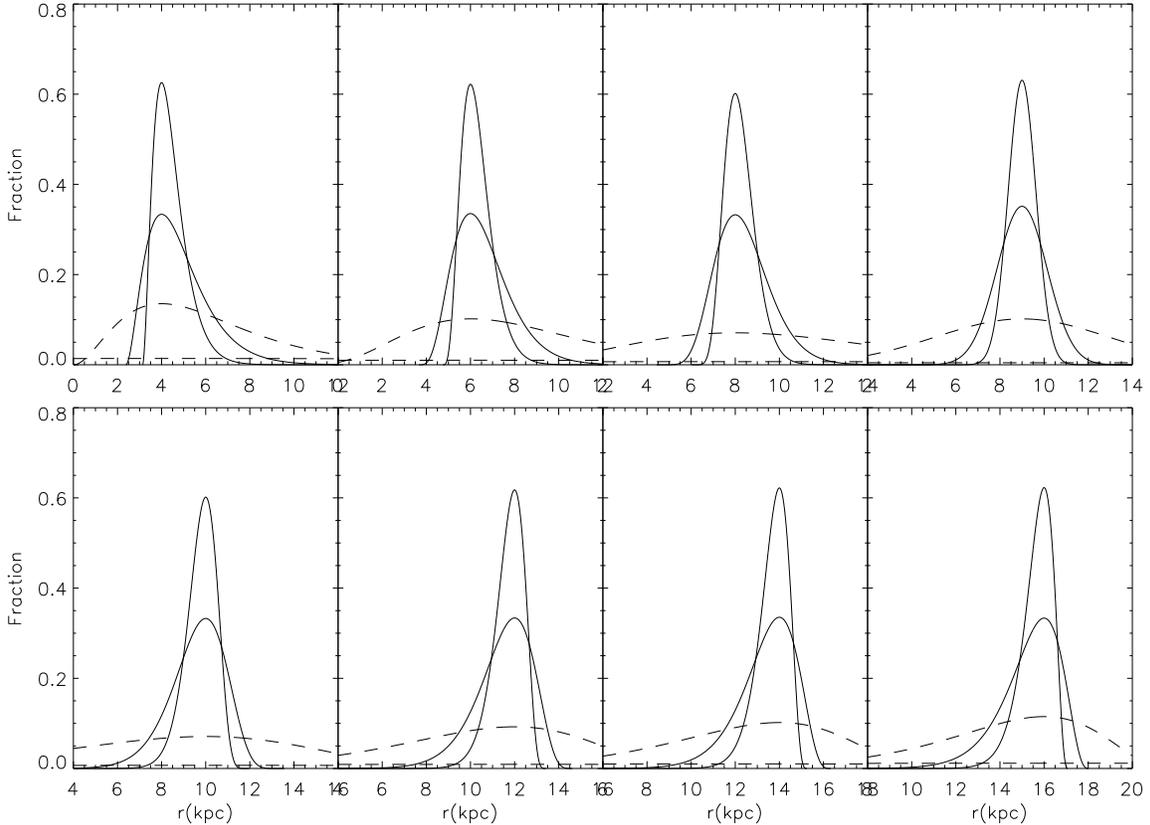}
\caption{Gama distribution for radial stellar migration for stars with birth radii $R=4,6,8,9~kpc$ (first row from left to right), and $R=10,12,14,16~kpc$ (second row from left to right). The solid line and dash-dotted line represent the standard and extreme condition, respectively. For the same kind of line, taller lines are for 2~Gyr-old stars and lower lines are for 6~Gyr-old stars. The normal distribution at all radii in the pure normal distribution is the same as that at $R=9~kpc$.}
\label{distribution}
\end{figure*}

The distribution functions we adopt for radial stellar migration are normal distribution with stellar-age-dependent $\sigma$ and Gama distribution with stellar-age-dependent shape and scale parameters. For the normal distribution,
\begin{equation}
\centering
f_{S\!M}=f_{N}(r)=\frac{1}{\sigma\sqrt{2\pi}} \, \exp{\left[-\frac{1}{2} \left(\frac{r-\mu}{\sigma}\right)^{2}\right]}
\end{equation}
where $\mu$ is the expectation corresponding to the maximum and the center of $f_{N}(r)$, and $\sigma$ is the variance representing the width or the shape of the distribution. We make the center at the birth radii of stars and $\sigma$ increase with the stellar age exponentially so that stars would stay around their birth radii with the highest probability and migrate further with time elapsing (see the panel for birth radius $R=9~kpc$ in Figure \ref{distribution}). The normal distributions for all radii are adopted to be same.

Following the hint of Figure 13 of \citet{SB02} which shows the distributions of radii of stars with different initial birth radii, we adopt Gama distribution $\Gamma(\alpha,\beta)$ which is a kind of asymmetrical distribution. The basic form of $\Gamma(\alpha,\beta)$ is:
\begin{equation}
f_{S\!M}=f_{G}(r)=\frac{1}{\beta^{\,\alpha}\,\Gamma(\alpha)} \, r^{\,\alpha-1} \, \exp{\left(-\frac{r}{\beta}\right)}~,\: r>0
\end{equation}
\begin{equation*}
\Gamma(\alpha)=\int\limits^{\infty}_{0} t^{\,\alpha-1} e^{-t} dt
\end{equation*}
where $\alpha$ is the shape parameter and $\beta$ is the scale parameter. Considering the shape of Gama distribution will become more symmetrical as the shape parameter $\alpha$ increasing and wider with the scale parameter $\beta$ getting bigger, we make $\alpha$ radius-dependent indicating stars born at different radii would have different migration tendencies or preferences, and $\beta$ would increase with the stellar age exponentially. At the same time, we also make the maximum of the Gama distributions correspond to the birth radii of stars. At $R=9~kpc$, where is a transition radius, we adopt the normal distribution as same as in the pure normal distribution introduced above.

For both the normal and Gama distributions, we adopt two scenarios for the variation of their profiles with stellar age: standard condition and extreme condition. In the extreme condition, stars would migrate much faster than in the standard condition, which can't be possible in reality. We just want to test what the effects would be if stars migrate under an extreme scenario. Table \ref{Gama} shows the values of $\sigma$ in normal distribution, and $\alpha$ and $\beta$ at all radii in Gama distribution in both standard and extreme conditions, for the transition radius $R=9~kpc$  the normal distribution as same as in the pure normal distribution is adopted. In Figure \ref{distribution} we show the profiles of Gama distribution at different radii for stellar age of 2~Gyr (taller line) and 6~Gyr (lower line), and the normal distribution at all radii is the same as that at $R=9~kpc$. The solid line and dash-dotted line represent the standard and extreme condition, respectively. Here we point out that for simplicity the migration probability adopted here is constant in time.

\begin{table}
\caption{Values of $\sigma$ in the normal distribution, $\alpha$ and $\beta$ at all radii in the Gama distribution in both standard and extreme conditions.}
\begin{center}
\begin{tabular}{cccc}
\hline \hline
\noalign{\smallskip}
Normal & & & \\
\hline
Radius &  & $\sigma$ & $\sigma$  \\
(kpc)  &  & standard  & extreme  \\
\noalign{\smallskip}
\hline
\noalign{\smallskip}
all R &  &  $exp(0.088t)-0.56$  &  $exp(0.75t)-0.56$  \\
\noalign{\smallskip}
\hline \hline
\noalign{\smallskip}
Gama & & & \\
\hline
Radius & $\alpha$ & $\beta$ & $\beta$  \\
(kpc)  &          & standard  & extreme  \\
\noalign{\smallskip}
\hline
\noalign{\smallskip}
4 & 3  &  $exp(0.071t)-0.72$  &  $exp(0.5t)-0.72$  \\
\noalign{\smallskip}
5 & 4  &  $exp(0.061t)-0.77$  &  $exp(0.5t)-0.77$  \\
\noalign{\smallskip}
6 & 5  &  $exp(0.054t)-0.80$  &  $exp(0.5t)-0.80$  \\
\noalign{\smallskip}
7 & 6  &  $exp(0.0495t)-0.82$  &  $exp(0.5t)-0.82$  \\
\noalign{\smallskip}
8 & 10  &  $exp(0.038t)-0.86$  &  $exp(0.5t)-0.86$  \\
\noalign{\smallskip}
9 & Normal  &    &    \\
\noalign{\smallskip}
10,11 & -10  &  $exp(0.038t)-0.86$  &  $exp(0.5t)-0.86$  \\
\noalign{\smallskip}
12,13 & -6  &  $exp(0.0495t)-0.82$  &  $exp(0.5t)-0.82$  \\
\noalign{\smallskip}
14,15 & -5  &  $exp(0.054t)-0.80$  &  $exp(0.5t)-0.80$  \\
\noalign{\smallskip}
16,17,18 & -4  &  $exp(0.061t)-0.77$  &  $exp(0.5t)-0.77$  \\
\noalign{\smallskip}
\hline
\end{tabular}
\label{Gama}
\tablecomments{t represents the age of star. In Gama distribution, for the transition radius $R=9~kpc$, the normal distribution as same as in the pure normal distribution is adopted. The negative value for $\alpha$ represents we adopt the symmetrical distribution of the function with the positive value of the $\alpha$.}
\end{center}
\end{table}

After migration, stars that die at radii different from their birth radii would eject their materials, including the newly synthesized materials and the ones taken from their birth radii, into the ISM at the positions they died, to some extent causing the mixture of material in large radius range in the disk. This may exert some effect on the Galactic chemical evolution.

\section{Contribution of stellar ejecta and nucleosynthesis prescriptions}

Stellar ejecta or yields are of significant importance in chemical evolution models. As the chemical enrichment of galaxies is basically due to stars, in spite of the light elements produced by primordial nucleosynthesis, the ejected materials of stars decide the amount of chemical elements in the ISM to a great extent. Here we adopt the stellar ejecta and yields calculated by models with convective overshoot \citep{M01, P98} for which the mass grouping is different from that of standard models \citep[e.g.][]{WW86, C92}, with lower critical masses. Here we only briefly show the definition of different mass groups which are classified according to the dominant physical processes governing their evolution and fate. For low-mass stars the upper mass limit is about $1.6\sim1.7~M_{\odot}$ in models with convective overshoot, and about $6~M_{\odot}$ for intermediate-mass stars, both depending on the chemical composition; quasi-massive stars will explode as ``electron capture'' supernovas, the mass range of which is about $\ge \sim6~to~8~M_{\odot}$; and massive stars are in the mass range $8\sim120~M_{\odot}$. (See \citet{P98} for more details. Here we changed their upper mass limit of intermediate-mass stars from $5~M_{\odot}$ to $6~M_{\odot}$, in consideration of the detailed explanation in their article.)

For low- and intermediate-mass stars ($m < 6~M_{\odot}$) we adopt the yields by \citet{M01} which calculates the stellar evolution tracks up to the TP-AGB and III dredge-up phase, and the ejecta of massive and quasi-massive stars ($6\sim120~M_{\odot}$) by \citet{P98}, where the mass loss by stellar wind has been considered for massive stars, are adopted here. These yields and ejecta are metallicity dependent and calculated basing on the same stellar models with the same series of stellar tracks for the whole range of stellar masses, so they are of good coherence and homogeneity of basic physical prescriptions. For type Ia SNe we adopt the ejecta $E_{SNeIa}$ of the W7 model of \citet{I99}. We assume each single star expels its ejecta all at once at the end of its life, while the binaries with the right properties to give rise to SNe Ia eject in two steps as introduced in Section 2.5, and the ejected material is immediately mixed in the ISM which remains always homogeneous.

\section{Observational data}

For the Galactic abundance distribution of iron we use the following sets of data: \citet{A02a, A02b, A02c, A04}, \citet{Lu03, Lu06}, \citet{LL11}, and \citet{L07, L08}, who analyzed the Galactic Cepheids. For the abundance of oxygen we adopt the data sets: \citet{A02a, A02b, A02c, A04}, \citet{L07}, and \citet{Lu03, Lu06}, who analyzed the Galactic Cepheids; \citet{S83}, \citet{E05}, \citet{VE96}, \citet{R97}, and \citet{R06}, who studied the Galactic HII regions; \citet{SR97}, \citet{G98}, and \citet{DC04}, who analyzed the Galactic OB stars. Considering the scatter of the data, we divided the data into 6 bins as functions of galactocentric distance, and calculated the mean abundance value and star-number-weighted radial position for each bin, to see the trend of the data more clearly. The mean value trends are represented by dark orange lines with filled circles in Figure \ref{Ogradient} and \ref{Fegradient}.

The data for the surface gas density profile along the Galactic disk are adopted from \citet{D93}. Concerning the metallicity distribution of the G-dwarf stars in the solar neighborhood, we use the data of \citet{RM96}, \citet{J00} and \citet{K02}.

\section{Results}

In this section we present the results of our models.

\begin{figure}
\centering
\includegraphics[width=3.4in, angle=0]{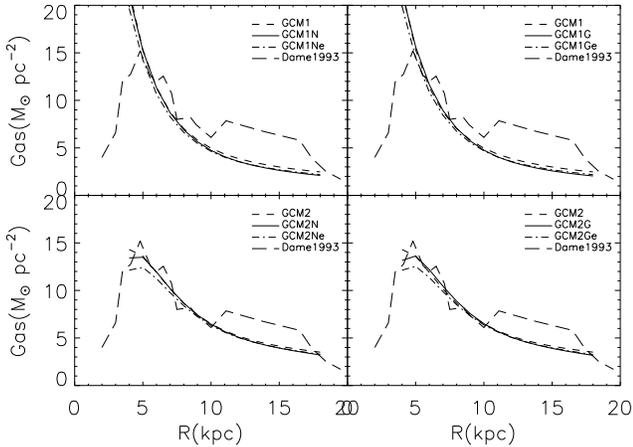}
\caption{Surface mass density distribution of gas along the Galactic radius. Solid lines and dash-dotted lines in each panel represent the results of models with stellar migration of standard and extreme condition, respectively. Dashed lines represent the results of basic models without stellar migration. Observational data from \citet{D93} are represented by long-dashed lines.}
\label{ISM}
\end{figure}

\begin{figure}
\centering
\includegraphics[width=3.4in, angle=0]{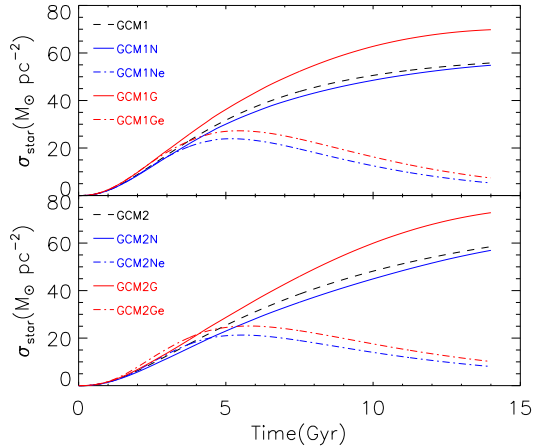}
\caption{Surface stellar mass density in the solar neighbourhood evolving with time. The solid lines and dash-dotted lines in each panel represent the results of models with stellar migration of standard and extreme condition, respectively. The colors of the lines, blue and red, represent the results of models adopting normal and Gama distributions for the stellar migration, respectively. The black dashed lines represent the results of basic models without stellar migration.}
\label{sigStar}
\end{figure}

In Figure \ref{ISM} we show the surface mass density of gas produced by our models. Solid lines and dash-dotted lines represent the results of models with stellar migration of standard and extreme condition, respectively. Dashed lines show the results of basic models without stellar migration. Long-dashed lines show the observational data from \citet{D93}. Comparing to basic models, there is less gas in regions where stars are lost by migration, since the materials ejected by these stars are also lost. As in the extreme condition stars migrate radially very fast, the gas produced by models with extreme stellar migration are less than that produced by models with standard stellar migration, especially in the inner disk where more stars are born. However, the influence of the radial stellar migration on surface mass density of gas is small. Figure \ref{sigStar} shows the surface stellar mass density in the solar neighbourhood evolving with time. For the standard condition, as more stars migrate out than into the solar vicinity, the stars produced by the models with normal-distribution stellar migration are less than that produced by basic models. But the stars produced by the models adopting Gama-distribution stellar migration are more than that produced by the basic models. This results from that more stars born in the inner disk would migrate to the solar neighborhood under the Gama-distribution radial stellar migration. However, the current surface stellar mass density in the solar neighborhood produced by all the models with standard stellar migration can roughly fit the observational values, such as $63~M_{\odot} \, pc^{-2}$ by \citet{J10} and $61\pm6~M_{\odot} \, pc^{-2}$ by \citet{HF00} and \citet{HF04}. Since in those papers only the part under vertical height $z=1.1~kpc$ above the disk plane was taken into consideration, these values should be higher for the whole disk column. Thus the results calculated by the models with Gama-distribution radial stellar migration of standard condition should be closer to the real value. However, under the extreme stellar migration, large amount of stars migrate out of the solar vicinity very quickly. Thus the surface stellar mass density produced by models with extreme stellar migration show a trend of first increasing and then decreasing, and the current stars are much less than the observational value.

\begin{figure}
\centering
\includegraphics[width=3.4in, angle=0]{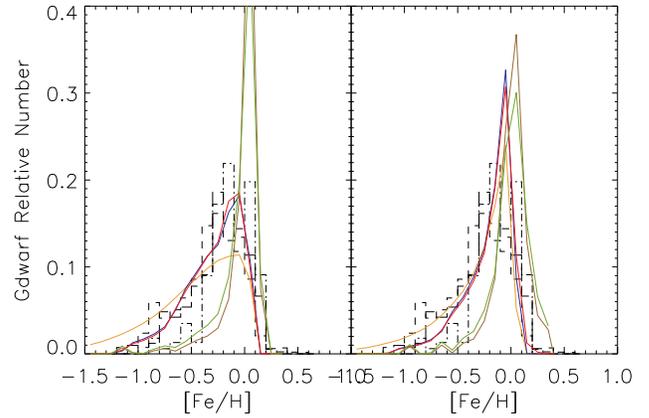}
\caption{G-dwarf metallicity relation in the solar neighborhood by models with constant disk formation time scale (left) and inside-out disk formation scenario (right). In each figure, the dark orange line represents the result from the basic model without stellar migration; the blue and red lines represent the results from the models including standard stellar migration with normal and Gama distribution respectively; the brown and green lines represent the results from the models including extreme stellar migration with normal and Gama distribution respectively. The dashed, dash-dotted and long-dashed histograms represent the observational data from \citet{K02}, \citet{J00} and \citet{RM96}, respectively.}
\label{Gdwarf}
\end{figure}
Considering the radial stellar migration, the G-dwarf metallicity distribution of the solar neighborhood should be different from the results from star-static models (see Figure \ref{Gdwarf}). As stars migrate radially, the solar neighborhood is not only assembled from local stars following a local age-metallicity relation, but also from stars originating from the inner (more metal-rich) and outer (more metal-poor) Galactic disk which have migrated to the current position with different timescales \citep{RR08b, SB09a}. Because of the higher density of stars in the inner disk, the migrated portion would favor metal-rich stars. As a consequence, the migrated metal-rich stars could compensate the metal-poor tail of the local chemical evolution, and the profiles of the G-dwarf metallicity distributions we derived become narrower and taller with less metal-poor stars, while the models with extreme condition really produce the extreme results, much more metal-rich stars. Thus, the stellar migration can be another solution to the ``G-dwarf problem''. However, the models adopting an inside-out disk formation, GCM2N and GCM2G, produce too many metal-rich stars in the solar neighbourhood, even though models adopting this kind of disk formation could have given out better results when do not consider the stellar migration. Thus radial stellar migration may become another mechanism for the disk formation competing with the inside-out disk formation, or they can coexist having similar effects.

\begin{figure}
\centering
\includegraphics[width=3.4in, angle=0]{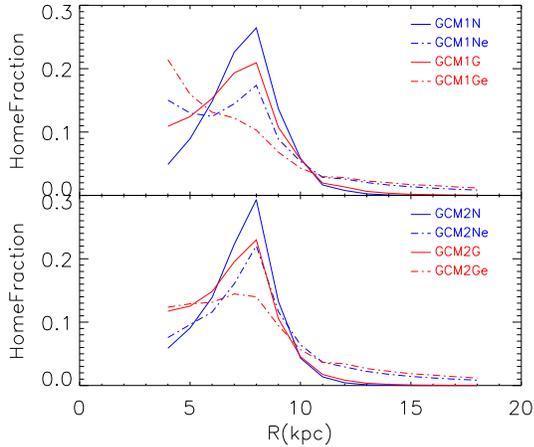}
\caption{Distribution of the birth radii of stars that appear at the solar neighbourhood at present. Solid and dash-dotted lines in each panel represent the results of models with stellar migration of standard and extreme condition, respectively. The colors of the lines, blue and red, represent the results of models adopting normal and Gama distributions for stellar migration, respectively.}
\label{Birthradii}
\end{figure}

\begin{figure}
\centering
\includegraphics[width=3.3in, angle=0]{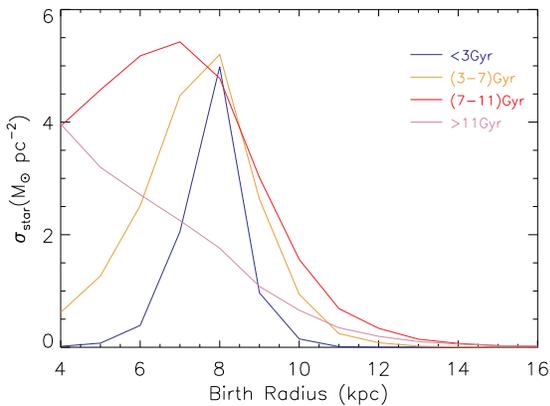}
\caption{Distributions of the birth radii of stars in different ages that appear at the solar neighbourhood at present by model GCM2G. Blue line is for stars younger than 3~Gyr; red line for stars of 3-7~Gyr; dark orange line for stars of 7-11~Gyr; and lilac line for stars older than 11~Gyr.}
\label{ABR}
\end{figure}

\begin{figure}
\centering
\includegraphics[width=3.4in, angle=0]{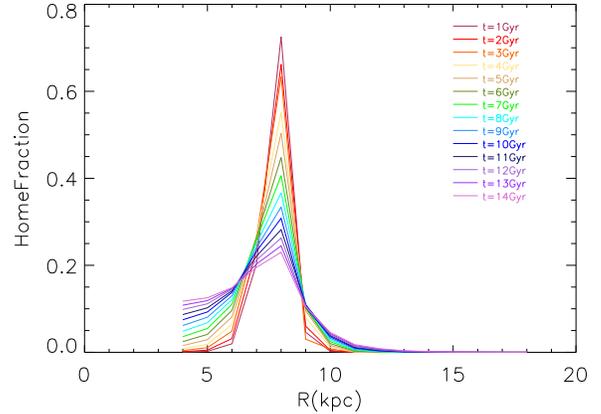}
\caption{Time-dependent variation of the birth radii distributions for stars in the solar vicinity by model GCM2G, for the Galactic evolution time from $t=1~Gyr$ to $t=14~Gyr$.}
\label{HRT}
\end{figure}

In Figure \ref{Birthradii} we show the distributions of birth radii of the stars that appear at the solar neighbourhood at present. As we already know, the stars in the solar neighbourhood could have been born in a large range of radius and migrated to their present positions. While most stars come from the local area, a large part of the stars come from inner disk, as a consequence of the higher density of stars there, which is more obvious for the models adopting Gama distribution for stellar migration in which stars in inner disk have a preference to migrate outwards. These phenomenon are more obvious for the extreme condition. For our best model GCM2G, we show the distribution of the birth radii of stars in different ages that appear at the solar neighbourhood at present in Figure \ref{ABR}. We can see that with the age increasing, the birth radii of stars cover larger range of the disk, and more stars come from inner disk, which is consistent with the stellar migration theory. We also show the time-dependent variation of the birth radii distributions for stars in the solar vicinity from the Galactic evolution time $t=1~Gyr$ to $t=14~Gyr$ in Figure \ref{HRT}. As the time goes, stars migrate for longer time, and more stars born at farther places come to the solar vicinity. Particularly for the stars from inner disk, we think that after about $t=4~Gyr$ they significantly enrich the stellar abundance in the solar vicinity, and this is also an explanation for the change of the G-dwarf metallicity relation caused by stellar migration which is shown in Figure \ref{Gdwarf}.

\begin{figure*}
\centering
\includegraphics[width=16.0cm, angle=0]{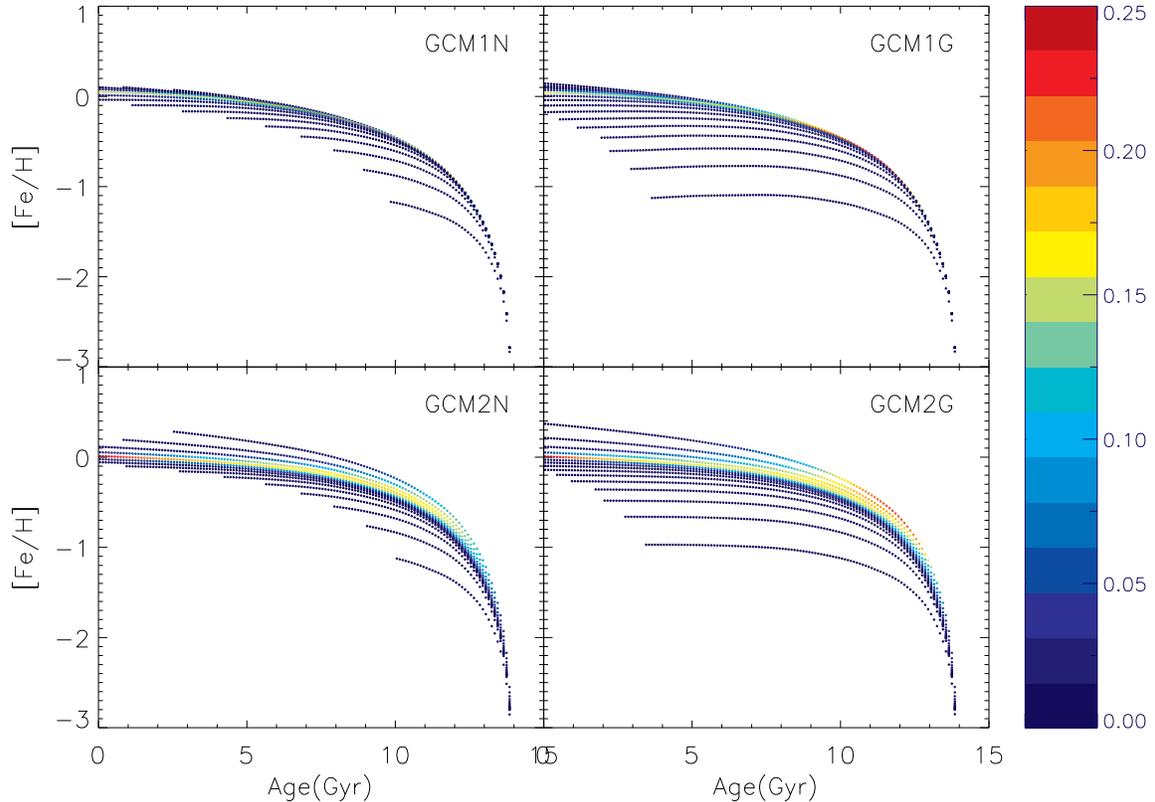}
\caption{Age-metallicity distribution in the solar neighbourhood. In each panel from the top to the bottom, the lines represent the stars migrated to the solar neighbourhood from their birth radii of $r=4~kpc$ to $18~kpc$, respectively, and the relative proportion are shown by the colors increasing from blue to red.}
\label{AMR}
\end{figure*}

The age-metallicity relation (AMR) of stars in the solar neighbourhood is an important factor when considering stellar migration. As a consequence of radial stellar migration, the solar neighbourhood consists of stars of different ages and metallicities coming from different radii, and they scatter the classic AMR to an age-metallicity distribution (AMD), as roughly shown in Figure \ref{AMR}. The lines in each panel, from the top to the bottom, represent the contributions of stars migrating from their birth radii of $r=4~kpc$ to $18~kpc$ respectively, and the relative proportion are shown by the colors (increasing from blue to red). As what we have expected, the AMR are scattered by the radial stellar migration, while a slope in the AMD is preserved and stars from the same birth radius still follow a classic local AMR. This phenomenon is also found by \citet{MCM12}.

\begin{figure*}
\centering
\includegraphics[width=16.0cm, angle=0]{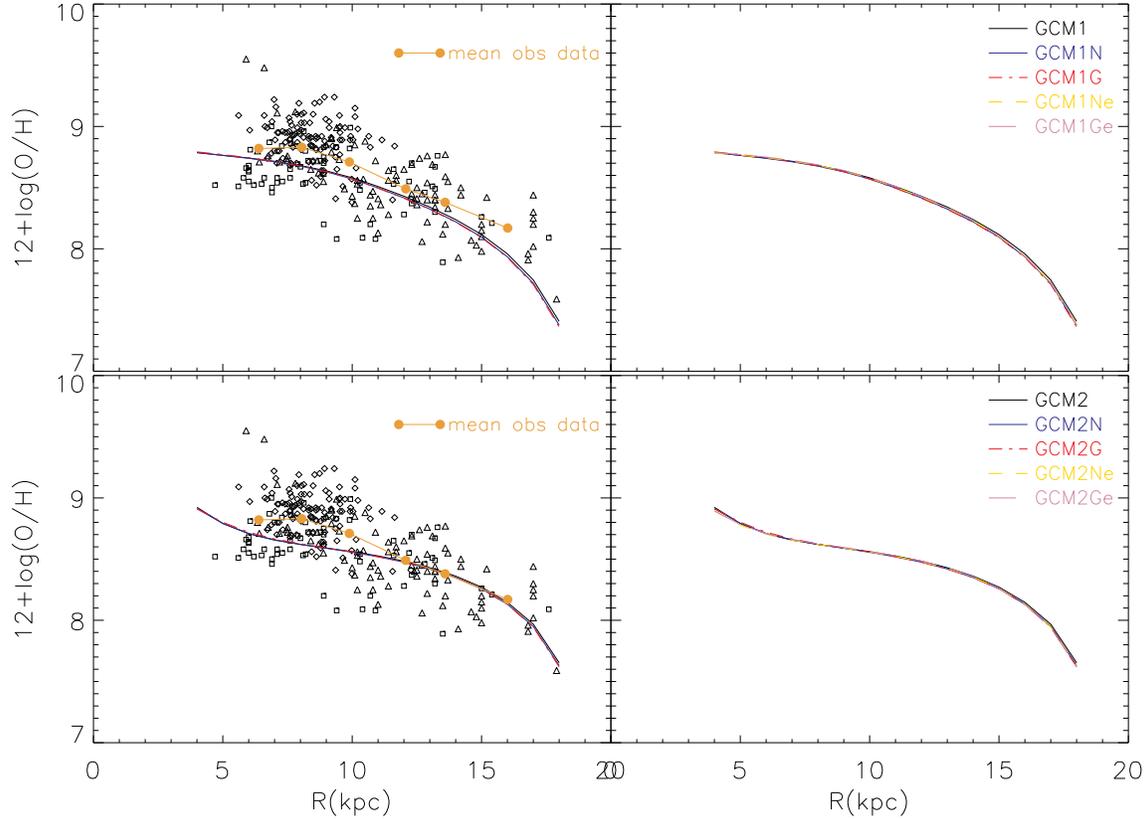}
\caption{Oxygen abundance along the disk obtained from models and observation. Panels on the first and second row are for models with constant disk formation time scale and inside-out disk formation scenario, respectively. Panels on the left column show the observational data and the results of models with standard radial stellar migration. The rhombuses, triangles and squares represent the observational oxygen abundance of Cepheids, HII regions and OB stars, respectively. The dark orange line with filled circles shows the mean value of the observational data. Panels on the right column only show the results of models, adding the results of models with extreme radial stellar migration, so that we can compare the results of different models clearly. In each panel, the black solid line represents the result of basic model without stellar migration; the blue solid line and the red dash-dotted line represent the results of models including standard stellar migration with normal and Gama distribution, respectively; the golden long-dashed line and the lilac dashed line represent the results of models including extreme stellar migration with normal and Gama distribution, respectively.}
\label{Ogradient}
\end{figure*}

\begin{figure*}
\centering
\includegraphics[width=16.0cm, angle=0]{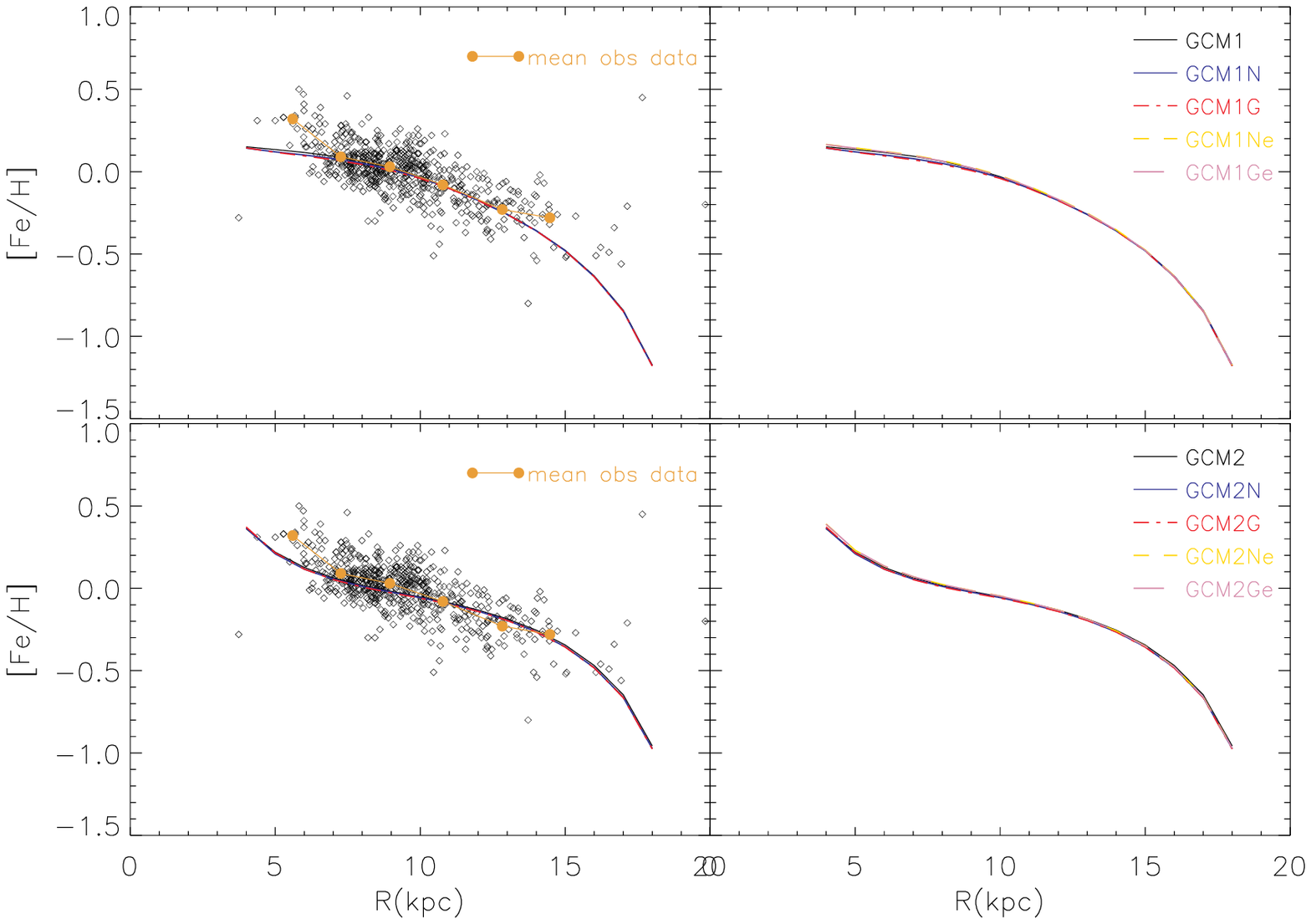}
\caption{Iron abundance along the disk obtained from models and observation. Panels on the first and second row are for models with constant disk formation time scale and inside-out disk formation scenario, respectively. Panels on the left column show the observational data and the results of models with standard radial stellar migration. The dark orange line with filled circles shows the mean value of the observational data. Panels on the right column only show the results of models, adding the results of models with extreme radial stellar migration, so that we can compare the results from different models clearly.} The types of the lines are of the same meaning in figure \ref{Ogradient}.
\label{Fegradient}
\end{figure*}

Figure \ref{Ogradient} and \ref{Fegradient} show the comparison among abundance distributions from models and observation. The panels on the first and second row are for models with constant disk formation time scale and inside-out disk formation scenario, respectively. From the left column, we can see that our models can reproduce the observational oxygen abundance and [Fe/H] distributions along the Galactic disk on the whole. In the right column we only compare the results from models. The abundance distributions in each panel, produced by the models adopting the same Galactic disk formation scenario, differ very little, almost coincide with each other. Being inconsistent with previous predictions, looking from the whole radius range, the distributions of chemical element abundances along the disk have not flattened after radial stellar migration, even in the extreme condition of radial stellar migration the abundance gradients have also not changed much. The specific values of the abundance gradient in the solar vicinity calculated by models are shown in Table \ref{gradientvalue}. These gradients are a little flatter than the typical values which are obtained by fitting the whole data set along the disk, such as for the iron gradient: -0.070 by \citet{L07}; -0.052 by \citet{L08}; -0.051 by \citet{P09}; -0.062 by \citet{LL11}, and for the oxygen gradient: -0.044 by \citet{E05}; -0.06 by \citet{R06}; -0.065 by \citet{L07}; -0.055 by \citet{LL11}. However, if only focus on the local gradient in the solar vicinity, we can find the gradient values for iron: -0.029 by \citet{A02a}; -0.03 by \citet{Lu03}; -0.044 by \citet{A04}; -0.042 by \citet{LL11}, which are consistent with our model results.

\begin{figure}
\centering
\includegraphics[width=3.5in]{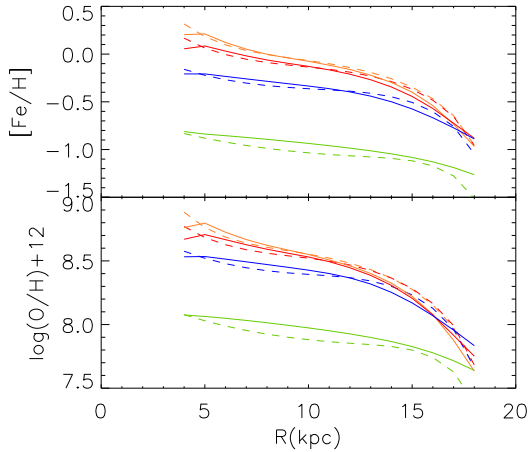}
\caption{Distributions of the mean iron (top) and oxygen (bottom) abundance of stars in different age groups calculated by model GCM2 (dashed lines) and GCM2G (solid lines). Orange lines represent stars younger than 3~Gyr; red lines represent stars of 3-7~Gyr; blue lines represent stars of 7-11~Gyr; green lines represent stars older than 11~Gyr.}
\label{SAG}
\end{figure}

\begin{figure}
\centering
\includegraphics[width=3.5in]{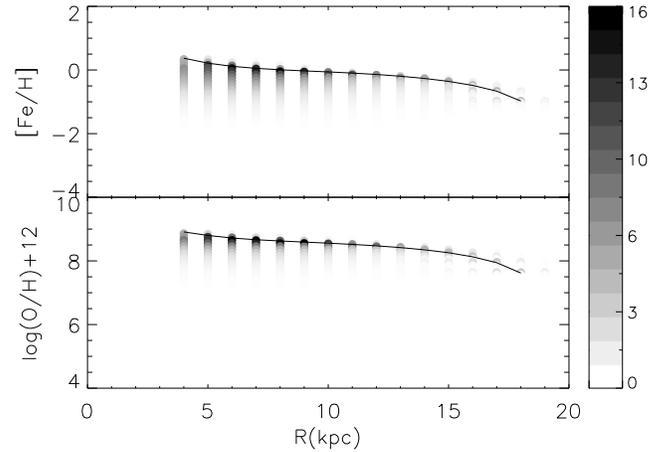}
\caption{Galactic iron (top) and oxygen (bottom) abundance distributions of stars by model GCM2G. The amount of star with a certain abundance value is shown by shade. As a comparison, the black solid lines show the abundance distributions of the ISM.}
\label{Stargradient}
\end{figure}

Then we may infer that the radial stellar migration can not lead to significant flattening of the chemical abundance gradients. This is not difficult to understand. As the important chemical contributors are massive stars, which have a short lifetime, they contribute mostly around their places of birth. Those migrated stars, born from far away having big difference in metallicity with the local stars, only contribute a small fraction to the local ISM and cannot change the chemical abundance of the local ISM essentially; and the chemical contributions by stars from nearby differ not much enough to change the abundance gradients significantly. However, as the migration is a motion of stars, its influence on the stellar chemical abundance is more significant. It can flatten the radial gradient of the mean chemical abundance of stars. As in Figure \ref{SAG}, we show the radial distributions of the mean iron and oxygen abundance of stars in four age groups by model GCM2 and GCM2G. Since older stars have migrated for more time, the influences of radial stellar migration on them are more significant, making their chemical abundances higher and abundance gradients flatter in the solar neighborhood, which is consistent with the findings of \citet{MCM12}. However, in the outer disk, as the disk there has formed much later and stars have a tendency to migrate inward under Gama-distribution migration, the chemical enrichment is slower and the mean stellar abundance gets lower for younger stars. Thus young stars have steeper abundance gradients in model GCM2G than in model GCM2. Besides, the most significant effect of the migrated stars on the chemical abundance at a position is they can scatter the abundance of stars there from a relatively concentrated value to a range. As shown in Figure \ref{Stargradient}, the shade shows the amount of star with a certain abundance value, which is calculated by model GCM2G as an example, and as a comparison the solid lines show the chemical abundance of the ISM. We can see that although the abundances of stars spread in a range, the abundance values which most stars possess at each radius still coincide with that of the ISM. This, from another way, suggests that the radial stellar migration cannot change the abundance gradient of the ISM essentially. Moreover, as the gradients here are not very steep, the effect of scatter becomes more significant.

\begin{figure}
\centering
\includegraphics[width=3.4in]{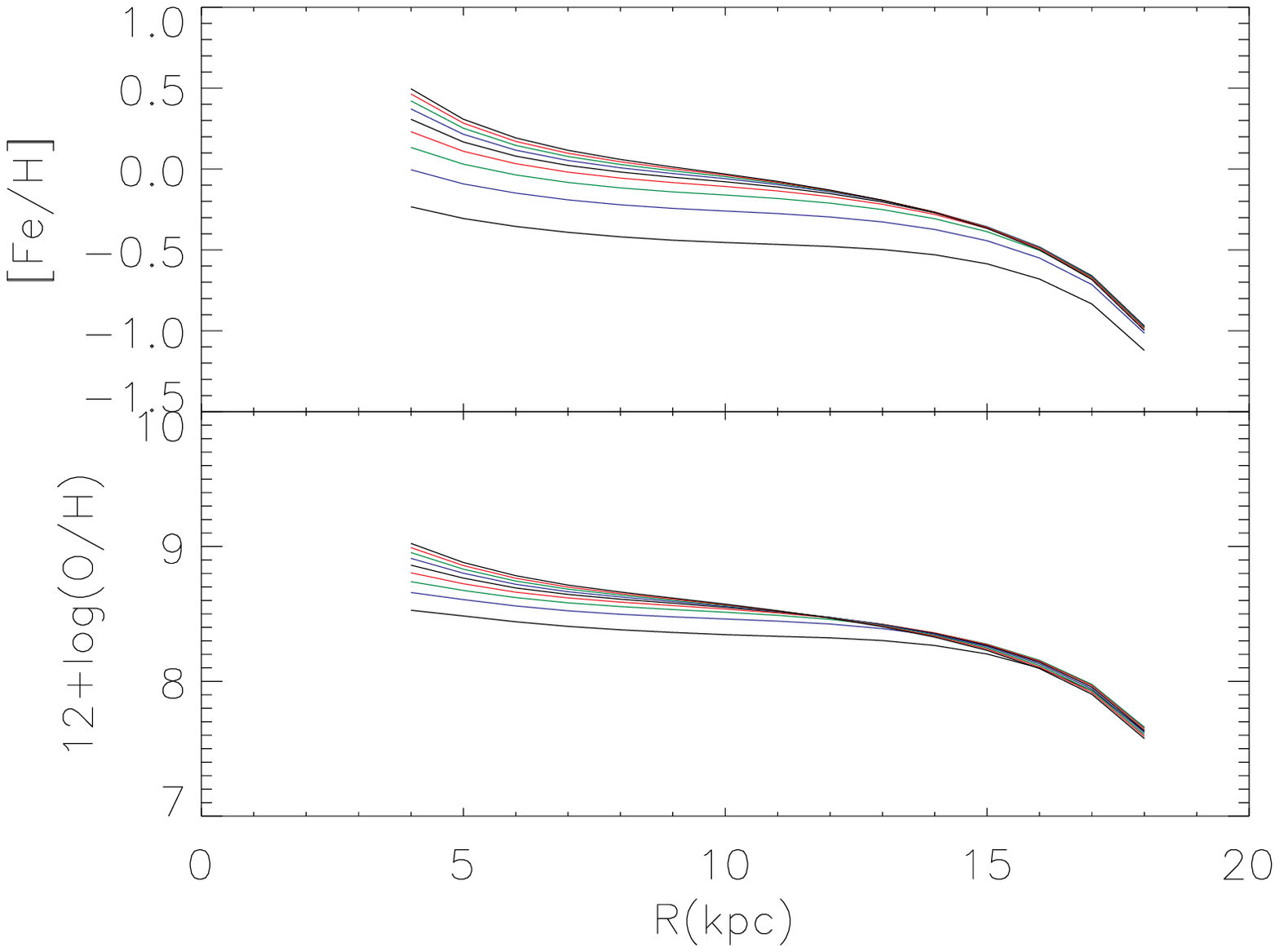}
\caption{Galactic radial iron (top) and oxygen (bottom) abundance distribution calculated by model GCM2G, the lines from the bottom to the top represent the abundance distributions from $t=4~Gyr$ to $t=20~Gyr$, respectively.}
\label{Gradients-time}
\end{figure}

We also predict that the abundance gradients of the Milky Way, such as the gradients of oxygen and iron, will not flatten under the influence of radial stellar migration in the future. On the contrary, while the gradients have kept steepening for the past 13.7~Gyr, they will continue steepening in the future, but with slower speed, as shown in Figure \ref{Gradients-time}.

\begin{table}
\caption{Abundance gradients of iron and oxygen in the solar vicinity.}
\begin{center}
\begin{tabular}{ccc}
\hline \hline
\noalign{\smallskip}
Model &    d[Fe/H]/dR    &  d[log(O/H)+12]/dR   \\
      & $(dex~kpc^{-1})$ &  $(dex~kpc^{-1})$    \\
\noalign{\smallskip}
\hline
\noalign{\smallskip}
GCM1 & -0.036 & -0.039  \\
\noalign{\smallskip}
GCM1N & -0.033 & -0.041  \\
\noalign{\smallskip}
GCM1Ne & -0.036 & -0.041  \\
\noalign{\smallskip}
GCM1G & -0.033 & -0.042  \\
\noalign{\smallskip}
GCM1Ge & -0.036 & -0.041  \\
\noalign{\smallskip}
GCM2 & -0.041 & -0.033  \\
\noalign{\smallskip}
GCM2N & -0.040 & -0.035  \\
\noalign{\smallskip}
GCM2Ne & -0.042 & -0.033  \\
\noalign{\smallskip}
GCM2G & -0.040 & -0.036  \\
\noalign{\smallskip}
GCM2Ge & -0.042 & -0.033  \\
\noalign{\smallskip}
\hline
\end{tabular}
\label{gradientvalue}
\end{center}
\end{table}

\section{Summary and conclusions}

The aim of this paper is to study the influence of radial stellar migration on the chemical evolution of the Milky Way, especially for the Galactic abundance gradients. We did our research mainly based on two detailed Galactic chemical evolution models: one adopts a constant disk formation time scale for all the Galactocentric radii; the other adopts an inside-out disk formation in which the disk formation time scale increase with radius. According to the basic point, we found the optimal forms for the radial gas flow and values for the parameters in the two models respectively to make them reproduce the observational data well. Then we added the radial stellar migration to these two models, where we used the distribution functions to model the radial stellar migration, and tested its effects in standard and extreme conditions, respectively. From the new models we derived the elemental abundance distributions along the Galactic radius and other important information.

Our main conclusions are summarized as follows:

\begin{itemize}

\item When stars migrate radially on the Galactic disk, the solar neighborhood is not only assembled from local stars, but also from stars originating from the inner and outer Galactic disk. As there are more stars in the inner disk which are of higher metallicity, more metal-rich stars would migrate to the solar neighbourhood. As a consequence, the profiles of the G-dwarf metallicity distributions produced by our models considering the stellar migration are narrower and taller with smaller metal-poor tails than those produced by the models without stellar migration. Thus, the radial stellar migration can be another solution to the ``G-dwarf problem''.

\item As model GCM2 adopting an inside-out disk formation can produce taller and narrower G-dwarf metallicity distribution than model GCM1 adopting a constant disk formation time scale, it seems that the radial stellar migration and the inside-out disk formation have some similar effects in building up the G-dwarf metallicity distribution. So whether the Galactic disk has formed inside-out or how it has formed in an inside-out way is still a question.

\item As stars could migrate radially, the stars in the solar neighbourhood could have been born in a large range of radius on the Galactic disk. Since older stars have migrated for more time, their birth radii cover larger range of the disk than younger stars, with larger fraction of them coming from inner disk. On the whole, under our assumption of the stellar migration, most stars in the solar neighbourhood come from the local area, while a large fraction comes from the inner disk, and fewer come from the outer disk.

\item The classic AMR is scattered by the stellar migration to an age-metallicity distribution, while a slope in the distribution is preserved and stars from the same birth radius still follow a classic local AMR.

\item We compared the results of models including radial stellar migration with those obtained from models without migration. Looking from the whole radius range, we find that the distributions of chemical element abundances along the disk haven't changed much after radial stellar migration. Different from previous prediction, we infer that the radial stellar migration cannot change the chemical abundance gradients of the ISM essentially. However, it can flatten the radial gradients of the mean chemical abundance of stars. Since older stars have migrated for more time, they possess flatter abundance gradients than younger stars.

\item The most significant effect of radial stellar migration on the chemical abundance at a position is it scatters the abundance of stars there from a relatively concentrated value to a range, while keep the dominant abundance value consistent with the abundance of the ISM.

\item According to the model results, we predict that the chemical abundance gradients of the Milky Way, such as the gradients of oxygen and iron, will continue steepening in the future, but with slower speed.

\end{itemize}

Finally, we should point out that we just use two basic chemical evolution models as a foundation to study the effects of radial stellar migration. We are not sure our models can produce the best results to fit all the observational data. Only according to our models in this paper, we also can not decide what kind of the combination of disk formation scenario, gas flow, stellar migration and the related parameters can make one model reproduce the observational data best. But we believe that our conclusions can, to a certain extent, tell the influences of radial stellar migration on the Galactic chemical evolution, and more further detailed research is still needed.

\begin{acknowledgements}
We thank the anonymous referee for helpful comments. We also thank F. Matteucci and Y. Liang for their advice for the construction of the model, and R. Chang and X. Fan for many useful discussions. This work is supported by the National Natural Science Foundation of China under grant No. 11233004.
\end{acknowledgements}

\end{document}